# Development of a Cloud-Based Payroll Management System

**Isaac Odun-Ayo and Adeyemi Aina**

Department of Computer and Information Sciences, Covenant University, Ota, Ogun State, Nigeria

isaac.odun-ayo@covenantuniversity.edu.ng     adeyemi.aina@stu.cu.edu.ng

**Abstract**

Cloud computing is continually evolving, enhancing hardware technologies, improving software and enhancing business processes. A payroll management system deployed on the Cloud harnesses on-demand of delivery of computational power and database storage using cloud computing technologies. This project aims to develop and deploy a cloud-based payroll management system. The objectives of this study are: to carry out a study on the existing cloud-based payroll management system, to design a payroll data model for calculating basic salary and enables retrieval of payroll history when needed from the database, to develop and deploy a payroll management system, on the Cloud that generates earning statements, filling the gap between security infrastructure and optimal system performance harnessing cloud computing technologies. The focus was on the design, implementation and deployment, using UML diagrams to illustrate the payroll application and Google App Engine for deployment. The system analysis in comparison of a conventional payroll system and the cloud-based system is endless in terms of speed, processing power, storage capacity, universalization and pricing. The cloud-based payroll has an infinite number of advantages; all conventional payroll system is rendered obsolete as it mends all the cons.

***Keywords***: payroll, cloud computing

## 1. Introduction

Cloud computing plays an increasingly important role in the operations of organizations of all sizes and all industries around the world. Cloud computing is a broad-ranging set of technologies and business practices. Cloud computing is a model for distributing process and capacity resources on request (Shinder, 2019). Cloud computing technologies are used with different software and applications today, as the payroll management system. A payroll management system is an organized list or records of employees and their salaries; a payroll management system is often used to generate the total amount of money that organizations pay to its employees (Moss, 2016). A payroll management system is a software or application which has a purpose for organizing, arranges and maintains all the tasks of employee payment which includes various deductions that need to be collated on the employee's payroll. Given the pay rate, the tasks may incorporate monitoring days or hour worked, deduction of taxes, adding compensations and reasonings, printing and conveying checks and paying business taxes to the government (Grande et al., 2011). A cloud-based payroll management system; this introduces the use of Cloud computing, Cloud computing is continually changing, making new hardware technologies, improving software and enhancing business processes. The historical backdrop of computing is right around a consistent stream of advances(Sullivan, 2009), of which the payroll management system is deployed on the Cloud can harness on-demand of delivery of computational power and database storage using cloud computing technologies (Varia & Mathew, 2014). Cloud-based payroll management offers better approaches to offer types of assistance while fundamentally modifying the cost structure hidden those services, the advantages a cloud-based payroll system over a conventional system are speed, accuracy, capacity, cost reduction, While lowering cost is a top priority, scalability, mobility, connectivity, and business agility have stepped to the forefront for decision-makers, these will be discussed in detail as we move on.

There are existing cloud-based payroll management system, used by large enterprises, in comparison these applications have a rather complicated user interface, there is so much more that can be done, which makes this cloud-based payroll management system more efficient, this project cannot be deployed on regular web hosting, because it is a Software as a Service(SaaS) whose scale cannot be determined, but the companies, so flexibility is essential. Deployment on google App Engine gives reports on the health of deployed application like error tracebacks, security testing and even seamless integrations like cloud scheduling and cloud SQL. Versioning the application and splitting traffic across different version enabling a vast majority to use this application without problems, deploying microservices which are called cloud functions which automatically manages the whole application on the Cloud.

The project will likewise fill in as a manual for researchers who might need to build more sophisticated systems using cloud technologies and develop updated versions to serve optimally. The Conventional Payroll management system has become an obsolete system of the employee information is stored locally, the speed of the computation depends on the processing capacity of the system's hardware, the processing speed, amount of RAM, the conventional Payroll management system is only as efficient as the system it is installed. The cloud-based payroll management system should be embraced as all the cons listed about the conventional system does not apply to a cloud-based payroll system.

## 2. Related Work

Bragg (2003) showed services provided by a payroll management system; this describes the payroll system process flows for explicit sorts of systems: outsourced payroll, a payroll application, and in-house manual payroll. Which includes: - set up of new employees, collect time card information, verify time card information, generate employee wages, enter employee changes and other processes.

Babic (2019) proposed different payroll data models for calculating basic wages and allowances of employees; this payroll data models include: - salaries as agreed by employment contract per year, net salaries with specific amounts deducted for taxes (pay rate per time), salaries paid monthly.

Mahajan et al., (2015) illustrated a comparison of conventional and cloud-based payroll management system which grouped into features like speed, efficiency, processing of data, ease of use and cost, which describes all these features of the conventional payroll management system depending on or limited to the specification of the system.

Rao & Rao (2014) presented an overview of deployment of a payroll management system on the Cloud, the payroll management system which handles the financial aspects of employee's or worker's wages, allowances, bonuses, deductions, taxes, gross pay and net pay. And the generation of pay rates and slips for a particular period. Using cloud computing technologies of which the application works remotely deployed on the Cloud.

Ghal et al. (2018) described Cloud resources and technologies contributing to a cloud-based payroll management system which includes content delivery network, automatic scaling also known as autoscaling, load balancing; they are all managed by cloud functions.

Grande et al. (2011) described what cloud computing takes to fill the gap between security infrastructure and optimal system performance harnessing cloud computing technologies. Cloud computing provides multi-factor authentication, data and information on the Cloud are encrypted, also serverless execution environment for building and connection to cloud services

## 3. Materials and Method

### 3.1 Design a Payroll Data Model

A payroll data model allows quick calculation an employees' salary; a payroll data model is used whether running a small or large organization, there needs to be some payroll solution, to understand all the data required for such a payroll management system(Babic, 2019), we will walk through the related payroll data model. There will always be differences in regulations, company policies and government taxes. The payroll data model in three subject areas

- Salaries, as agreed by an employment contract, are per year.
- Net salaries (i.e. with specific amounts deducted for taxes.) are paid to employees.
- Salaries are paid monthly.

### 3.2 Deploying a Payroll Management System

Using Google Cloud Platform Engine, it is a platform provided as a service, called platform as a service(PaaS), it eliminates the need to manage a server or operate hardware and other infrastructure, also using Cloud Storage which can store terabytes of data, Cloud Datastore which is a memory for serving data to applications with low latency and also Cloud SQL relational database that can run on persistent disks in terabytes in size(Ghal et al., 2018)

### 3.3 Optimal performance and security on the Cloud

Google's multi-layered approach to ensure that security is built-in by design at the data level. Google App Engine provides services that help organizations meet key developer needs with user authentication, versioning, and robust security tools like Cloud Security Scanner to eliminate threats to data security, e.g. hacks, viruses, denial-of-service (DoS)(Bevan & Banks, 2018)

Optimal performance of a system using cloud tools like Automatic scaling to meet any demand, Load balancing, version in for the application to migrate or split traffic on a network, multi-factor user authentication, logging, task queues, monitoring and other support packages.

## 4. Result and Discussion

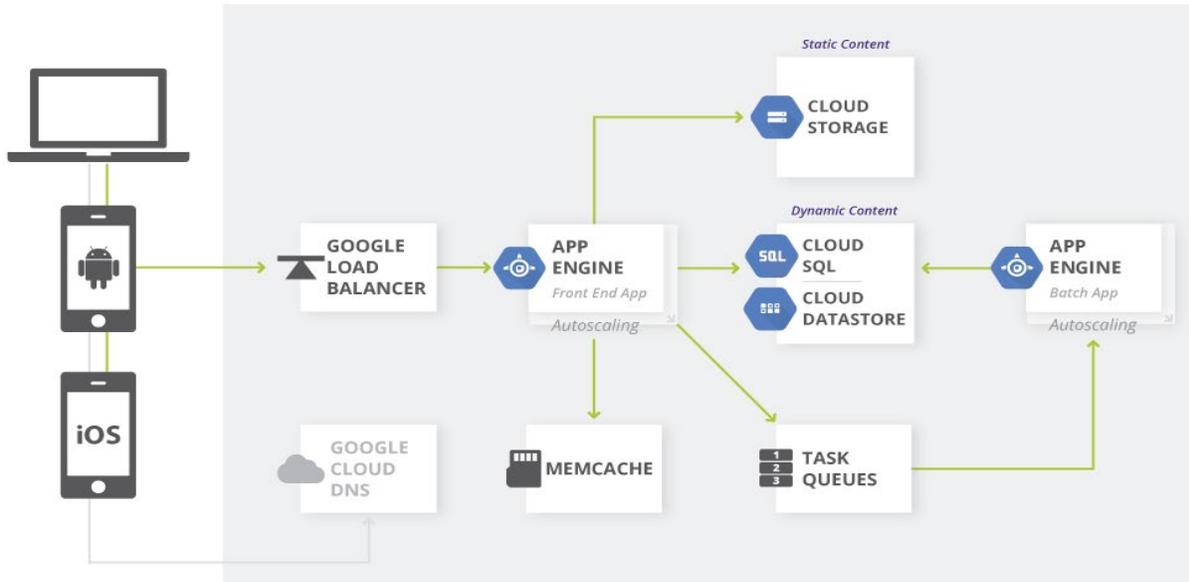

Figure 1: Deployment Architecture for Google App Engine

From Figure 1, Google App Engine is a flexible, zero ops platform for building highly available applications. Among the primary services and structures available are Google Load Balancer, which manages the load balancing of the payroll management system; Front End App, responsible for redirecting requests for appropriate services; Memcache, the cache memory shared between instances of Google App Engine, generating high speed in the availability of the information on the server, e.g. quick retrieval of payroll receipt or history; and Task Queues, a mechanism that provides redirection of long tasks (generating employee payroll) to back-end servers, making front-end servers free for new user requests. Also, Google App Engine also has static and dynamic storage solutions. The former provides the file storage service called Cloud Storage, whereas the latter provides relational database services such as Cloud SQL which can run on persistent disks over a terabyte in size, and non-relational NoSQL such as Cloud Datastore which is a low latency memory for serving data to applications.

| Employee Earning | | | |
|---|---|---|---|
| Description | Rate | hours | Current |
| Regular pay | N2500 | 45 | N112,500 |
| Gross Earnings | | | N112,500 |

| Employee Taxes Withheld | | Employer Taxes | |
|---|---|---|---|
| Employee Taxes | Current | Company Taxes | Current |
| Federal Income Tax | N11,250 | Medicare | 400 |
| Fees & Tolls | 250 | Insurance | 300 |
| State Income Tax | 250 | | |

Employee Contributions

Figure 2: Implementation of a payroll data model

From Figure 2, with designing a data model, there is no payroll data model generally applicable to every business. There are always differences in regulations, company policies and governments. That will require the model to be customized to cover the needs of a specific payroll. However, the principles laid out in this model should be relevant to most organizations. The payroll data model used is net salaries (i.e. with specific amounts deducted for taxes.) pay rate per time.

Table 1: Comparison of a conventional system and cloud-based System

| Features | Conventional Payroll System | A cloud-based payroll system |
|---|---|---|
| | Conventional Payroll system is enabled to store data while managing the data locally on a system | A cloud-based payroll system is a software application deployed on the Cloud using a cloud computing technology and provides on-demand database, storage and power for computation. |
| Speed | A conventional payroll system the speed depends on the processing capacity of the system's hardware, the processing speed, amount of RAM. | A cloud-based payroll system has unlimited speed despite the gravity of the computation, of which the Cloud provides delivery of computational power. |
| Efficiency | A conventional payroll system is efficient depending on the system's hardware or system installed on. | A cloud-based payroll system is more efficient despite the system's hardware as less time required |
| Processing of information or data | Processing data is slower, based on a single system processing power of a large amount of data. | Processing data is more rapid as the delivery of computational power with cloud technologies of even large volumes of data are dealt with |
| Ease of use | The conventional payroll system can only be used on specific techniques, of which the applications are installed on, if these systems are unavailable, there is no software to work. | The cloud-based payroll system possesses universality in which this application can be used on any network. |
| cost | The conventional payroll system, expenses are associated with the software incorporate preparing and program maintenance. Costs include fast with charges for other supplies | The cloud-based payroll system Expenses as associated pay-as-you-use pricing, of which are made only when the application is used, and the cloud service manages maintenance and security. |

From 1, we can see the comparison of a conventional payroll system to a cloud-based payroll system and how limited the conventional payroll system is, compared to the cloud-based payroll management system.

## 5. Conclusion

The cloud-based payroll management system, deployed on google cloud as software as a Service(SaaS), which manages the financial related aspects of employee's payment and age of pay slips for a period. The Conventional Payroll management system has become an obsolete system of the employee information is stored locally; the conventional Payroll management system is only as efficient as the system it is installed. The cloud-based payroll management system should be embraced as all the cons listed about the conventional payroll management system does not apply to a cloud-based payroll system. This project is subject to succeeding reviews and revaluation various areas, in the nearest future, new technology and features will be added to the existing components of the system for an optimal experience, so the payroll management system will keep evolving just as we have used latest technologies like cloud computing which helps in the delivery of power, database storage and security threats over a while.

## Acknowledgement

We acknowledge the support and sponsorship provided by Covenant University through the Centre for Research, Innovation, and Discovery (CUCRID). There are no conflicts of interest.